\documentclass[fleqn,usenatbib,twocolumn]{mnras}

\usepackage{lmodern}
\usepackage{bm}
\usepackage[T1]{fontenc}
\usepackage{newtxtext,newtxmath}
\usepackage{graphicx}
\usepackage{amsmath}
\usepackage{booktabs}
\usepackage{microtype}
\usepackage{xspace}

\newcommand{\msun}{\mathrm{M}_\odot}

\newcommand{\teff}{T_{\rm eff}}

\newcommand{\hii}{H\,{\sc ii}\xspace}

\newcommand{\Lspec}{\mathscr{L}}
\newcommand{\tform}{t_{\rm form}}
\newcommand{\mf}{m_{\rm f}}

\usepackage[dvipsnames]{xcolor}
\usepackage{soul}
\usepackage[normalem]{ulem}

\newcommand{\draftfigure}[2]{%
\IfFileExists{#1}{\includegraphics[width=\linewidth]{#1}}{%
\fbox{\begin{minipage}[c][0.30\textheight][c]{0.94\linewidth}\centering
Missing figure: \texttt{\detokenize{#1}}\\[0.7em]
#2
\end{minipage}}}}

\title[Slow Massive-Star Formation and ZAMS Deficit]{Slow Massive-Star Formation and the O-Star ZAMS Deficit}

\author[Padoan et al.]{%
Paolo Padoan,$^{1,2}$\thanks{E-mail: ppadoan@icc.ub.edu}
Samuel Shipkoff,$^{3}$
Mark Gieles,$^{2,4,5}$
Fabrice Martins$^{6}$
\\
$^{1}$Department of Physics and Astronomy, Dartmouth College, Hanover, NH 03755, USA\\
$^{2}$Institut de Ci\`encies del Cosmos, Universitat de Barcelona, Mart\'i i Franqu\`es 1, E-08028 Barcelona, Spain\\
$^{3}$Waterville Valley Academy, Waterville Valley, NH 03215, USA \\
$^{4}$ICREA, Pg. Llu\'{i}s Companys 23, E08010 Barcelona, Spain\\
$^{5}$Institut d'Estudis Espacials de Catalunya (IEEC), Edifici RDIT, Campus UPC, 08860 Castelldefels (Barcelona), Spain\\
$^{6}$LUPM, Universit\'e de Montpellier, CNRS, F-34095 Montpellier, France
}

\date{Accepted XXX. Received YYY; in original form ZZZ}
\pubyear{2026}

\begin{document}
\label{firstpage}
\pagerange{\pageref{firstpage}--\pageref{lastpage}}
\maketitle

\begin{abstract}
The scarcity of optically visible O stars near the theoretical zero-age
main sequence (ZAMS) has puzzled astronomers for over four decades. We
argue that it reflects prolonged accretion. If massive stars assemble
over Myr timescales, as the inertial-inflow model predicts, core-hydrogen
burning begins before accretion ends, and the stars become optically
visible at the end of accretion on a birthline displaced from the
standard ZAMS. We test this interpretation by comparing
the hot-side envelope of Galactic O stars in the Kiel diagram with
accreting stellar-evolution models, adopting a growth law
$t_{\rm form}=\tau_0(m_{\rm f}/60\,M_\odot)^\alpha$ with formation time
$t_{\rm form}$ and final mass $m_{\rm f}$. The resulting
birthlines reproduce both the displacement and the high-mass turnover of
the observed envelope, with best-fitting parameters
$(\alpha,\tau_0)=(0.52,4.3\,{\rm Myr})$, consistent with independent
constraints from the luminosity functions of compact H\,{\sc ii} regions
and OB stars. The two constraints have orthogonal degeneracy directions,
so combining them narrows the allowed parameter space, yielding
projected 90\% confidence intervals $0.43<\alpha<0.60$ and
$3.9<\tau_0/{\rm Myr}<5.0$. The inferred growth law agrees with
cloud-scale simulations and the inertial-inflow model, and implies that
mass assembly and main-sequence evolution overlap substantially in
massive stars, with broader implications for stellar evolution and
feedback.
\end{abstract}


\section{Introduction}
For over four decades, optically visible O stars have been known to be
scarce near the theoretical zero-age main sequence (ZAMS, the locus
conventionally associated with the establishment of stable core-H
burning; \citealt{Iben1965}), yet the physical origin of this deficit
remains unclear. \citet{Holgado2020} found a pronounced deficit near the ZAMS
for initial masses of approximately $30$--$70\,\msun$, while
\citet{Schootemeijer2021} found that Small Magellanic Cloud stars above
$\simeq20\,\msun$ are similarly rare during the first half of core-H
burning. A complementary discrepancy appears during the embedded phase:
compact \hii regions are far more abundant relative to optically
visible O stars than expected if they are freely expanding nebulae
around stars that have completed their formation, the classical
compact-\hii-region lifetime problem
\citep{WoodChurchwell1989,Churchwell2002,Hoare2007,Mottram2011}.
Thus, too few optically visible stars are found near the ZAMS, while
too many compact \hii regions are observed during the embedded phase.
Although generally treated separately, both discrepancies may measure
the duration of massive-star growth. \citet{PadoanGieles2026} used the
compact-\hii-region and OB-star luminosity functions as one clock of
that growth. Here we test whether the Galactic O-star ZAMS deficit can
provide a second, independent clock.

The ZAMS deficit can constrain the growth time because the stellar birthline need not coincide with the standard ZAMS. Here the \emph{standard ZAMS} is the ZAMS locus of non-accreting tracks initialized directly at a prescribed stellar mass, which is also their maximum mass. We define the \emph{birthline} as the locus at which accreting stars end their main growth phase; for massive stars, this also approximately marks their local optical emergence from the natal dense gas \citep{BeechMitalas1994,BernasconiMaeder1996}. The relation between the birthline and the ZAMS depends on the accretion and Kelvin--Helmholtz timescales, $t_{\rm acc}$ and $t_{\rm KH}$, respectively. For low- and intermediate-mass stars ($\lesssim10\,{\rm M}_\odot$), $t_{\rm acc}<t_{\rm KH}$, so accretion ends before sustained core-H burning begins. The birthline therefore lies away from the ZAMS and an optically
visible, nearly fixed-mass pre-main-sequence contraction connects the two \citep{Stahler1983,PallaStahler1990}. For massive stars, $t_{\rm KH}<t_{\rm acc}$, so core-H burning begins while accretion is still ongoing, eliminating the visible pre-main-sequence phase. If accretion then continues for an appreciable fraction of the
main-sequence lifetime, the star consumes part of its central hydrogen while still embedded and reaches a birthline displaced to lower effective temperature from the standard ZAMS. 

This possibility was demonstrated by the constant-accretion models of
\citet{BeechMitalas1994} and the mass-dependent models of
\citet{BernasconiMaeder1996}. Their Myr-scale formation times displaced
the massive-star birthline from the standard ZAMS and showed that
massive stars can already be burning hydrogen while still accreting and remaining embedded. Despite these early results, rapid formation became the
prevailing theoretical expectation. In the turbulent-core model,
massive stars form through the rapid collapse of preassembled massive
cores, on timescales of approximately $10^5$ yr
\citep{McKeeTan2002,McKeeTan2003}. Such a short growth phase allows little nuclear evolution before
accretion ends, leaving little time to displace the birthline from the
standard ZAMS, and is also shorter than the few $10^5$ yr duration
inferred from compact-\hii-region demographics
\citep{WoodChurchwell1989,Churchwell2002,Hoare2007,Mottram2011}.

A physically different picture is provided by the inertial-inflow model
(IIM) of \citet{Padoan2020}. In a supernova-driven simulation of a
representative Galactic environment, \citet{Padoan2020} find that the
initial collapsing core contains only a small fraction of the mass
eventually attained by the star; most of the mass is supplied
subsequently by parsec-scale converging motions already present in the
turbulent velocity field. Massive-star growth is therefore controlled
neither by the collapse of a preassembled massive core nor by
competitive accretion from a shared cluster reservoir, in which
protostars compete for gas within the common gravitational potential
\citep{BonnellEtAl2001,BonnellBate2006}. Its duration is instead set by
the turnover time of the turbulent flow on the scale from which the
mass is collected. Because only a fraction of the gas on that scale reaches a given star,
massive stars require large supplying regions and grow over Myr
timescales. The IIM predicts that the formation time increases
approximately as $\tform\propto\mf^{1/2}$, where $\mf$ is the stellar mass assembled by the end of formation before accounting for wind losses, with a normalization that depends on the parent-cloud
environment. Long formation times have also emerged from independent
cloud-scale simulations \citep[e.g.][]{Grudic2022,Kaalva+26}.

The recent joint modelling of compact-\hii-region and optically
revealed OB-star luminosity functions has provided an independent
observational test of this prediction, yielding a Myr-scale growth law
with $t_{\rm form}\propto m_{\rm f}^{1/2}$
\citep{PadoanGieles2026}. Here we test the same growth law with a
second clock, using accreting stellar-evolution models computed with
the Modules for Experiments in Stellar Astrophysics
(MESA; \citealt{Paxton2011,Paxton2013,Paxton2015,Paxton2018,Paxton2019}).
We define the birthline from the models' end-of-accretion points in the
Kiel diagram (stellar surface gravity versus effective temperature)
and compare it with the full observed envelope of Galactic O stars in
the same diagram. By the \emph{observed envelope}, we mean the hot-side
boundary of the stellar distribution. We show below that the Kiel-diagram
and luminosity-function constraints have transverse degeneracy
directions in the $(\alpha,\tau_0)$ plane, so their combination
substantially tightens the constraints on the growth law.

%Our primary diagnostic is the spectroscopic HRD (sHRD). The classical HRD plots bolometric luminosity, $L_{\rm bol}$, and it therefore depends on distance, extinction, bolometric corrections, multiplicity, and the adopted temperature scale. The sHRD instead uses
%\begin{equation}
%\log\left(\frac{\Lspec}{\Lspec_\odot}\right)
%      =4\log\left(\frac{\teff}{T_{\rm eff,\odot}}\right)
%       -\log\left(\frac{g}{g_\odot}\right),
%\label{eq:shrd}
%\end{equation}
%which can be constructed directly from atmosphere-model fits \citep{LangerKudritzki2014,Castro2014}.  
%Recent work shows that the same population can look closer to the ZAMS in the classical HR diagram than in the sHRD.  In Cygnus OB2, \citet{Berlanas2020} found that the sHRD reveals an evident gap above $\simeq32\,\msun$ even when the HR diagram appears less decisive.  In Carina, \citet{Berlanas2025} found stars close to or on the ZAMS in the HR diagram, but recovered an absence of stars close to the ZAMS above $\simeq25\,\msun$ in the sHRD.  This tension motivates using the fitted $\teff$ and $g$ as the primary test of whether a massive star is actually on the standard ZAMS.

\section{Massive-star birthlines and the Galactic O-star distribution}
\label{sec:models}

To construct the birthlines of accreting stars, we first specify
the formation time as a function of $m_{\rm f}$. In the
IIM \citep{Padoan2020}, this time is set by the turnover time of the
turbulent flow supplying the star, so its normalization depends on the
parent-cloud environment. For approximately virialized star-forming clouds, the expected environmental dependence can be written as \citep{Gieles2025}:
\begin{equation}
\tform(\mf) \simeq
1\,{\rm Myr}
\left(\frac{\mf}{100\,\msun}\right)^{1/2}
\left(\frac{\Sigma_{\rm g}}{10^3\,\msun\,{\rm pc}^{-2}}\right)^{-3/4}
\left(\frac{M_{\rm g}}{10^5\,\msun}\right)^{1/4} ,
\label{eq:tform_env}
\end{equation}
where $\Sigma_{\rm g}$ and $M_{\rm g}$ are the mean gas surface density and gas mass of the parent cloud, respectively. The model also predicts an approximate upper limit, $m_{\rm f}\lesssim 2.5\times10^{-3}M_{\rm g}$ \citep{Gieles2025}. Equation~(\ref{eq:tform_env})
therefore implies a distribution of formation times across Galactic
star-forming environments rather than a universal normalization. In
the present analysis we do not model this environmental variation for
individual O stars, but instead seek a characteristic growth law for
the Galactic population. We parameterize it as
\begin{equation}
\tform(\mf)=\tau_0
\left(\frac{\mf}{60\,\msun}\right)^\alpha ,
\label{eq:tform_fit}
\end{equation}
where $\tau_0$ is the characteristic Galactic formation time at the
pivot mass $60\,\msun$ and $\alpha$ determines the mass dependence.
The luminosity-function analysis of \citet{PadoanGieles2026} yielded
$\tau_0=4.15^{+0.90}_{-0.63}\,{\rm Myr}$ and
$\alpha=0.48\pm0.11$; here we explore the ranges
$1\leq\tau_0/{\rm Myr}\leq7$ and $0\leq\alpha\leq1.5$.

\begin{figure*}
\centering
\includegraphics[width=\textwidth]{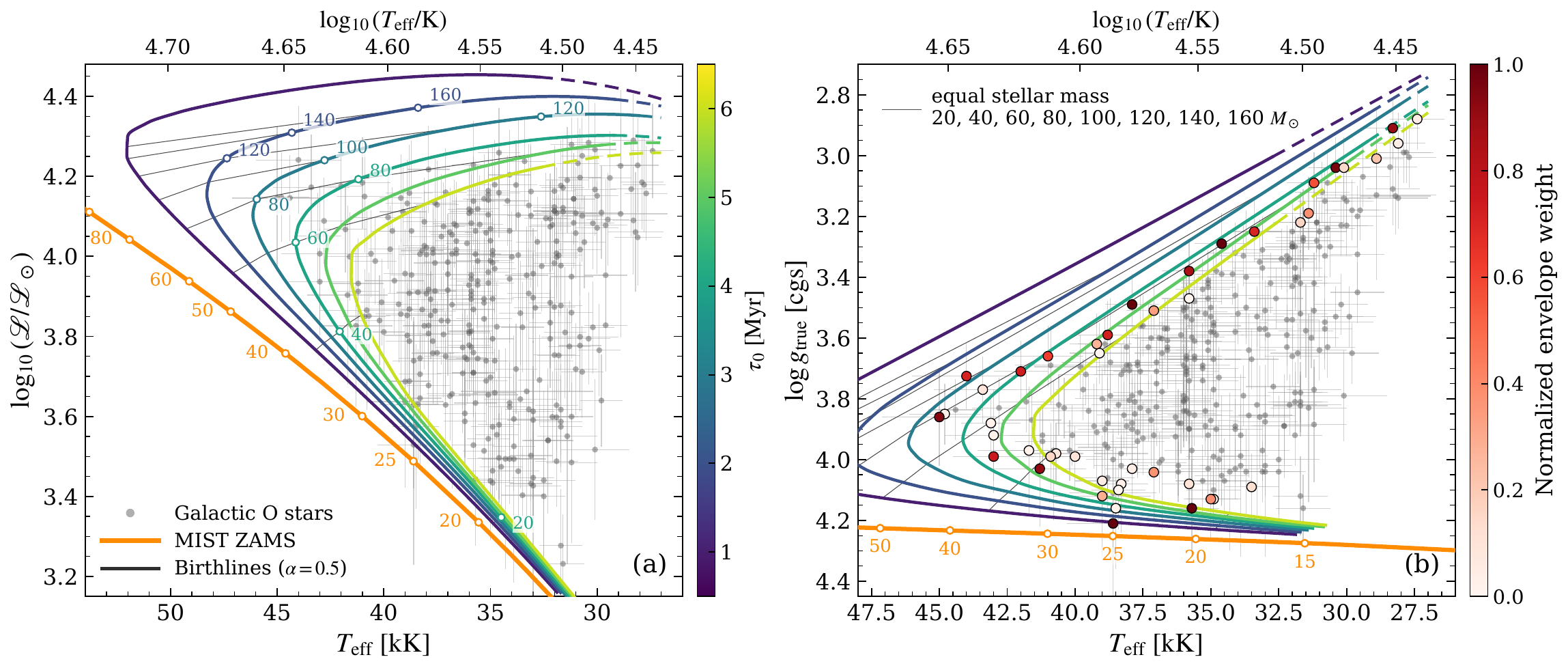}
\caption{
Comparison of the Galactic O-star distribution with the predicted massive-star birthlines.
The grey points show the Galactic O-star sample in the spectroscopic Hertzsprung--Russell diagram (left) and the Kiel diagram (right). The orange curve is the standard non-rotating MIST ZAMS. Coloured curves are the birthlines, defined by the end-of-accretion points of accreting models obeying $t_{\rm form}=\tau_0(m_{\rm f}/60\,M_\odot)^\alpha$, for $\alpha=0.5$ and $\tau_0=1$--$6\,{\rm Myr}$ as indicated by the colour scale. Selected stellar masses are labelled along the ZAMS and several birthlines in solar-mass units, while thin black curves connect points of equal current stellar mass across the birthline family. Dashed segments are geometric continuations of the calculated birthlines to lower effective temperature (see text). In the Kiel diagram, the highlighted stars trace the observed hot-side envelope used in the fit, with colour indicating their normalized statistical weight.
}
\label{fig:birthline}
\end{figure*}

We compute non-rotating, solar-metallicity accreting models with
MESA (version 25.12.1), using the Ledoux criterion and a semiconvective
efficiency $\alpha_{\rm sc}=0.1$, as in MIST \citep{Choi2016}.
H-burning core-boundary mixing is treated with exponential
overshooting whose efficiency increases with current stellar mass,
guided by empirical main-sequence constraints
\citep{Brott2011,deBurgos2025,Sciarini2026} and multidimensional
calculations \citep{Baraffe2023}. In MESA's notation, $f$ sets the
exponential decay length beyond the convective boundary and $f_0$ the
interior reference point; we adopt full-strength values $f=0.035$ and
$f_0=0.008$, with MESA's native mass ramp increasing $f$ from zero at
$2\,\msun$ to full strength at $15\,\msun$. This corresponds
approximately to a step-overshoot parameter
$\alpha_{\rm ov}\simeq0.3$, comparable to values inferred for massive
stars \citep{Brott2011,Sciarini2026, Lechien2026}\footnote{We also computed accreting models without overshooting
for the same combinations of $\alpha$ and $\tau_0$.
The resulting birthlines differ only slightly from the
fiducial ones, leaving the inferred growth-law constraints
essentially unchanged.}. The high-mass calculations employ
MLT++ \citep{Paxton2013} to control radiation-dominated superadiabatic
layers.

Each model begins with the same pre-main-sequence seed model of mass
$m_{\rm seed}=2\,\msun$, and grows at the constant rate
\begin{equation}
\dot m_{\rm acc}(\mf)
=
\frac{\mf-m_{\rm seed}}{\tform(\mf)} ,
\label{eq:mdot}
\end{equation}
with $\tform(\mf)$ given by equation~(\ref{eq:tform_fit}).
Thus, for a fixed pair $(\alpha,\tau_0)$, $\mf$ determines both the duration and the rate of accretion. Stellar winds remain active during growth, so the actual stellar mass
at the end of accretion is smaller than $\mf$.

Our native MESA grid samples four growth-law exponents,
$\alpha=0,\ 0.5,\ 1$, and $1.5$, and six discrete values of
$\tau_0$ for each exponent: $\tau_0=2$--$7\,{\rm Myr}$ for
$\alpha=0$ and $\tau_0=1$--$6\,{\rm Myr}$ for the other three
exponents. These combinations define 24 distinct growth laws. For each growth law, we compute 17 or 18 accreting tracks with different values of $\mf$. Their end-of-accretion points provide a discrete sampling of the corresponding birthline, which we construct by interpolation in the Kiel diagram. In total, the 24 birthlines are based on 411 individual accreting tracks. We then construct a dense grid in $(\alpha,\tau_0)$ by interpolating between these native birthlines. 

We compare the models with a spectroscopic sample of Galactic O stars drawn from the ALS III parent population
\citep{PantaleoniGonzalez2025} used for the luminosity-function
analysis of \citet{PadoanGieles2026}. We use the IACOB samples of \citet{Holgado2020,Holgado2025}, supplemented with ALS III members from the Cygnus OB2 and Carina studies of
\citet{Berlanas2020,Berlanas2025}. For duplicate stars, we adopt the Berlanas parameters in those regions and otherwise give precedence to \citet{Holgado2025}. These studies already exclude double-lined spectroscopic binaries (SB2), whose composite spectra complicate the determination of reliable single-star parameters. We additionally exclude stars flagged as single-lined spectroscopic binaries (SB1) or suspected SB2 systems, together with four other stars with independent
indications of binarity or dynamical interaction.\footnote{
ALS~5039 is a confirmed SB1 \citep{TriguerosPaez2021};
HD~93128 \citep{Sota2014} and HD~93161~B \citep{Naze2005}
are suspected binaries; and HD~64568 is nitrogen enriched,
a runaway, and a plausible product of a dynamical interaction
between massive multiple systems
\citep{Markova2018,CarreteroCastrillo2025,MaizApellaniz2020}.} These exclusions reduce potential contamination by unresolved companions that can bias spectroscopic parameters and by stars whose evolution may have been altered by mass transfer or mergers. We also exclude the four remaining members of Trumpler 14 (three others are already excluded as known or suspected spectroscopic binaries), whose compact cluster environment may have a different formation-time normalization from the general Galactic population. The final sample contains 317 stars. For each star we use the published $\teff$, centrifugally corrected $\log g_{\rm true}$, and their uncertainties.

The left-hand panel of Fig.~\ref{fig:birthline} shows the adopted O-star sample in the spectroscopic Hertzsprung--Russell diagram (sHRD), whose ordinate $\Lspec\equiv\teff^4/g_{\rm true}\propto L/M$ is constructed entirely from spectroscopically inferred quantities and is independent of distance \citep{LangerKudritzki2014,Castro2014}. We use the sHRD only to
visualize the displacement from the standard ZAMS; the quantitative comparison is performed in the Kiel diagram shown in the right-hand panel. Figure~\ref{fig:birthline} also shows the standard non-rotating, solar-metallicity MIST ZAMS
\citep{Dotter2016,Choi2016} and, for illustration, birthlines for $\alpha=0.5$ and $\tau_0=1$--$6\,{\rm Myr}$. Increasing $\tau_0$ allows more core-hydrogen burning before accretion ends and therefore moves the birthline progressively away from the standard ZAMS. At high masses, the formation time becomes a substantial fraction of the main-sequence lifetime, so the effective temperature along the end-of-accretion sequence reaches a maximum and then decreases toward larger stellar masses, producing the turnover. Both the displacement and turnover are in qualitative agreement with the observed hot-side envelope. The equal-stellar-mass loci likewise show increasing displacement with $\tau_0$ at fixed stellar mass.

We refer to the portions of a birthline below and above the turnover in stellar mass as the \emph{lower-mass branch} and
\emph{higher-mass branch}, respectively. The corresponding portions of the observed hot-side envelope are the \emph{lower-mass} and \emph{higher-mass} envelopes, which together comprise the full observed boundary.

\section{Constraint from the Galactic O-star envelope in the Kiel diagram}
\label{sec:birthline}

We now use the observed hot-side envelope in the Kiel diagram to
constrain the growth-law parameters $(\alpha,\tau_0)$. We perform the
comparison in the Kiel diagram because $\teff$ and $\log g_{\rm true}$
are the two quantities inferred directly from the spectroscopic
analysis. A weighted observational envelope is constructed before
fitting any candidate birthline and is then held fixed throughout
the parameter grid for each choice of interval width. A fixed family of theoretical birthlines spanning a broad range of
growth-law parameters defines the local tangential and perpendicular
directions used to identify the observed envelope. For each reference
birthline, we select the star furthest towards the hot side within
overlapping intervals along the curve. Repeating the selection with
shifted interval grids and combining the selection frequencies across
the reference family yields the normalized weights $w_i$.
We repeat the selection for five choices of interval width
and adopt the middle choice as fiducial. The right-hand panel of
Fig.~\ref{fig:birthline} shows the fiducial envelope together with
the representative family of birthlines for $\alpha=1/2$ and
$\tau_0=1$--$6\,{\rm Myr}$. The full envelope-selection and weighting
procedure is described in Appendix~A of the supplementary
material.

For each $(\alpha,\tau_0)$, we measure the mismatch between the
corresponding birthline and the fixed observational envelope.
For each envelope star $i$, we define $d_i(\alpha,\tau_0)$
as its shortest distance to the birthline, with the temperature
and gravity differences normalized by that star's adopted
observational uncertainties. The overall mismatch is quantified by
\begin{equation}
Q(\alpha,\tau_0)
=
\sum_i w_i\,d_i^2(\alpha,\tau_0),
\label{eq:Q}
\end{equation}
where the envelope weights are normalized so that
$\sum_i w_i=1$. Thus, $Q^{1/2}$ is the weighted root-mean-square
separation between the observed envelope and the birthline in the
scaled Kiel diagram. We evaluate $Q$ over the full
$(\alpha,\tau_0)$ grid described in Section~\ref{sec:models}.

Because $Q$ measures the mismatch of a selected and weighted
boundary, its sampling distribution is not known analytically.
We therefore calibrate the confidence regions empirically using
mock O-star catalogues generated by resampling the observed
stellar distribution around fitted birthlines. For each mock
catalogue, we repeat the envelope selection, weighting, and
birthline fitting. Comparing the mismatch at the known input
parameters with the minimum obtained by refitting each mock
provides approximate confidence thresholds without invoking
a $\chi^2$ approximation. We calibrate the confidence regions
separately for the five selections of the observed envelope,
obtained using different interval widths, and report their
union. The mock construction and calibration are described
in Supplementary Appendix A.

\begin{figure}
    \centering
    \includegraphics[width=\columnwidth]{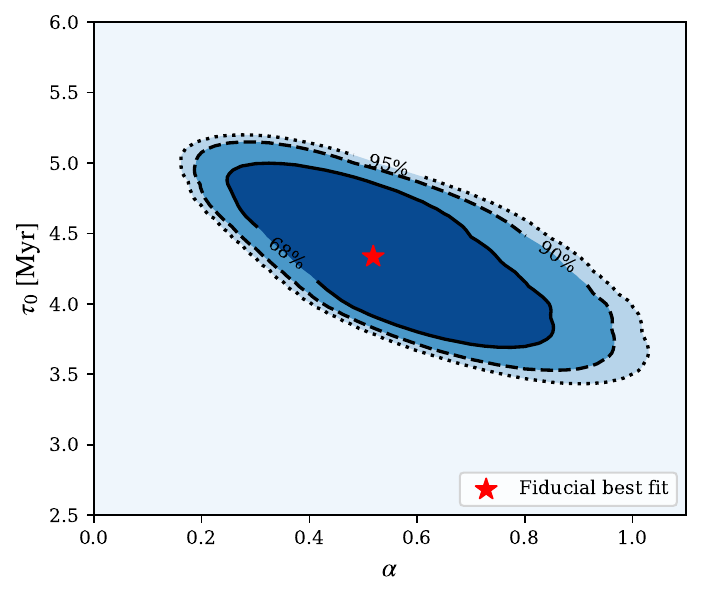}
    \caption{
    Constraint on the massive-star growth-law parameters $\alpha$ and
$\tau_0$ from the hot-side envelope of the Galactic O-star
distribution in the Kiel diagram. The dark, intermediate, and
light blue regions show the unions of the empirically calibrated
68\%, 90\%, and 95\% confidence regions, respectively, over the
admissible envelope-selection widths. The red star marks the
best-fitting parameters for the fiducial envelope selection,
$(\alpha,\tau_0)=(0.52,\,4.3\,{\rm Myr})$.
    }
    \label{fig:two_clocks}
\end{figure}

The resulting Kiel-envelope constraint is shown in
Fig.~\ref{fig:two_clocks}. For the fiducial envelope selection,
the best-fitting parameters are
$(\alpha,\tau_0)=(0.52,\,4.3\,{\rm Myr})$. The confidence regions
are elongated, with larger values of $\alpha$ corresponding to smaller
values of $\tau_0$, as expected if the formation time is constrained
most tightly by the better-defined higher-mass envelope. A steeper
mass dependence can then be compensated by a smaller normalization
while leaving the formation times of the most massive stars nearly
unchanged. We discuss the greater constraining power of the
higher-mass envelope further in Section~\ref{sec:discussion}. Thus,
the O-star envelope alone constrains both parameters of the growth
law, but leaves a substantial correlated uncertainty in the
$(\alpha,\tau_0)$ plane. In the following section, we combine this
constraint with the independent luminosity-function measurement of
the same growth law.

%As a limiting test, we compared the observed envelope with standard non-accreting main-sequence isochrones, corresponding to $\tau_0=0$. Their best fit occurs at an age of about $3\,{\rm Myr}$, but it cannot reproduce the full envelope simultaneously: matching the higher-mass portion displaces the turnover and lower-mass portion too far towards cooler temperatures. Thus, the limiting case $\tau_0=0$, in which the displacement from the ZAMS is produced entirely by post-formation evolution, cannot reproduce the shape of the observed envelope.

\begin{figure*}
    \centering
    \includegraphics[width=\textwidth]{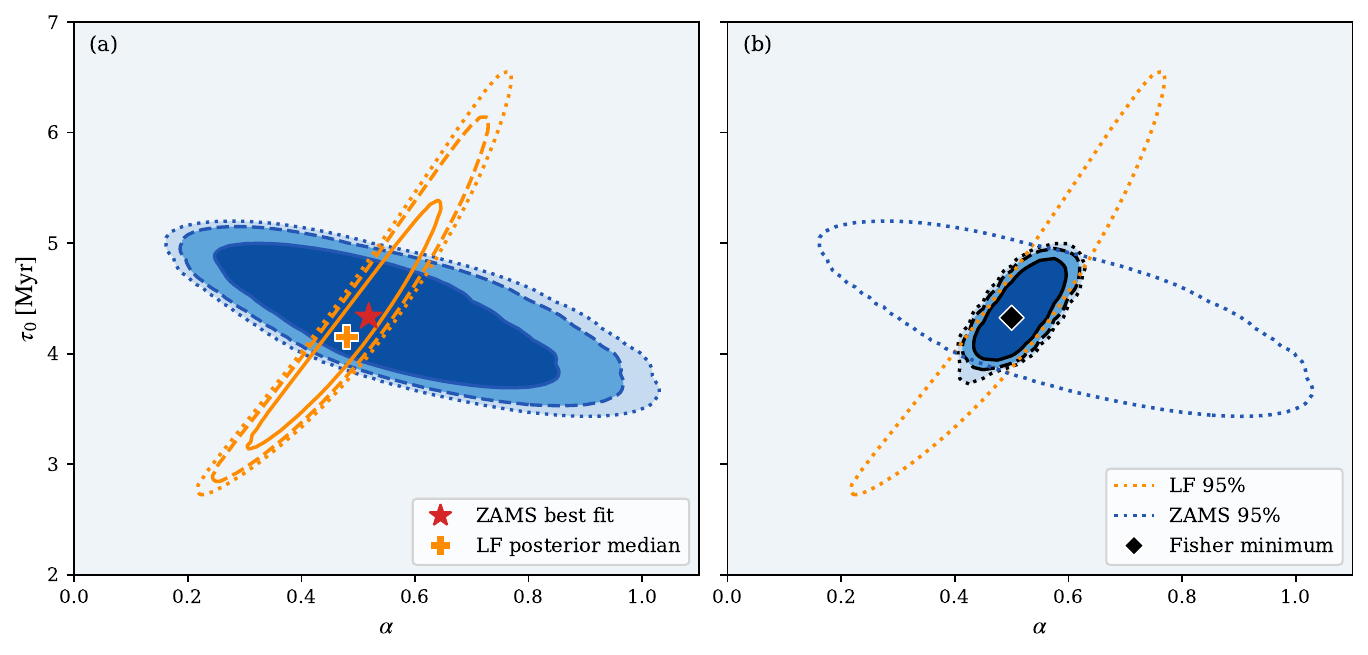}
    \caption{
    Constraints on the massive-star growth-law parameters $\alpha$ and $\tau_0$ from the Galactic O-star Kiel diagram and luminosity function. Left: the dark, intermediate, and light blue regions show the 68\%, 90\%, and 95\% confidence regions from the Kiel-diagram analysis, while the orange solid, dashed, and dotted contours show the corresponding luminosity-function constraints. The red star marks the best fit for the fiducial envelope selection, while the orange cross marks the luminosity-function posterior median. Right: the dark, intermediate, and light blue regions show the joint 68\%, 90\%, and 95\% confidence regions for $(\alpha,\tau_0)$ obtained by combining the Kiel-diagram and luminosity-function constraints. The dotted orange and blue contours show the 95\% luminosity-function and Kiel-diagram constraints alone, respectively, and the black diamond marks the Fisher minimum.
    }
    \label{fig:joint_constraints}
\end{figure*}

\section{Joint constraint on the growth law}
\label{sec:joint}

The luminosity-function analysis of \citet{PadoanGieles2026}
provides a complementary measurement of the same massive-star growth
law. As described in Section~2, the spectroscopic sample used here is
drawn from the same ALS~III parent population used in that analysis,
so the two measurements probe the same underlying Galactic population
and constrain the same parameters, $(\alpha,\tau_0)$, but through
different observables and modelling procedures. This common observational basis motivates a joint analysis, while
the different parameter degeneracies offer the possibility of
substantially improving the precision of the inferred growth law.

The left-hand panel of Fig.~\ref{fig:joint_constraints} compares the two constraints. 
The Kiel-envelope analysis favours $(\alpha,\tau_0)=(0.52,4.3\,{\rm Myr})$,
while the luminosity-function analysis gives posterior
median values of $(0.48,4.15\,{\rm Myr})$. Their confidence regions overlap substantially, showing that the two measurements are mutually
consistent. At the same time, the Kiel-envelope region is elongated along a direction in the $(\alpha,\tau_0)$ plane that is transverse to the luminosity-function constraint. Their intersection therefore localizes the growth-law parameters much more tightly than either measurement alone.

We combine the Kiel-envelope and luminosity-function constraints
using Fisher's method, as described in Supplementary Appendix B. 
For the fiducial envelope selection, the Fisher statistic
is minimized at
\begin{equation}
(\alpha,\tau_0)=(0.50,\,4.3\,{\rm Myr}),
\label{eq:joint_best}
\end{equation}
with projected 90\% intervals
\begin{equation}
0.43<\alpha<0.60,\qquad 3.9<\tau_0/{\rm Myr}<5.0.
\label{eq:joint_ranges}
\end{equation}

The right-hand panel of Fig.~\ref{fig:joint_constraints} shows the
joint 68, 90, and 95 per cent confidence regions, together with the
95 per cent Kiel-envelope and luminosity-function constraints for
reference. The joint region lies within the substantial overlap of the
two individual constraints but is much more compact because of their
transverse degeneracy directions. For comparison, the
luminosity-function constraint alone spans
$0.25<\alpha<0.73$ and
$2.9<\tau_0/{\rm Myr}<6.1$ at 90 per cent confidence.
Thus, combining the two observational clocks leaves the preferred
growth law close to the optima of both individual measurements while
substantially reducing its allowed parameter range.

\section{Discussion and conclusions}
\label{sec:discussion}

The principal result of this work does not require every star near the
observed envelope to be an unevolved single star. At high mass, main-sequence evolution after accretion ends proceeds
nearly parallel to the observed envelope, so differences in the time
elapsed since accretion ended predominantly move stars along the
boundary. At lower mass, the evolutionary tracks have a larger
component perpendicular to the envelope and therefore produce greater
scatter. Binary interaction can further broaden the lower-mass
envelope, because mass transfer and mergers may rejuvenate stars and
move them towards the ZAMS \citep[e.g.][]{deMink2014,Jin2026}.
The higher-mass envelope therefore provides a more sharply defined
constraint, although we fit the complete observed envelope and impose
no additional mass-dependent weighting.

A separate systematic concerns the spectroscopic determination of
surface gravity. The gravities used here were derived from
one-dimensional atmosphere models, whereas recent work indicates
that turbulent pressure can provide significant support in O-star
atmospheres and that neglecting it can lead to underestimated
gravities, particularly for giants and supergiants
\citep{GonzalezTora2025,Markova2025}. Correcting this effect would
primarily move the higher-mass envelope downward in the Kiel diagram,
with a much smaller effect on the lower-mass dwarf envelope. For the
geometry of our birthlines, such a shift would favour longer, rather
than shorter, formation times and therefore does not provide an
obvious route back to rapid massive-star formation. A quantitative
assessment will require a homogeneous spectroscopic reanalysis that
includes turbulent pressure self-consistently.

The numerical confidence regions remain conditional on the adopted
stellar models and observational inputs. More importantly for the
interpretation of $\tau_0$, our analysis represents the Galactic
O-star population with a single formation-time normalization, whereas
equation~(\ref{eq:tform_env}) predicts
$\tau_0\propto\Sigma_{\rm g}^{-3/4}M_{\rm g}^{1/4}$ at fixed stellar
mass. A distribution of formation times is therefore expected even
among ordinary Galactic molecular clouds: the observed dispersion in
surface density alone implies a factor-of-few range in $\tau_0$, while
the much higher surface densities of massive-cluster-forming
environments imply substantially shorter formation times. The value
$\tau_0\simeq4.3\,{\rm Myr}$ inferred here should therefore be
interpreted as an effective normalization for the Galactic O-star
population rather than as a universal value for individual
star-forming regions. Future population modelling should explicitly
include this environmental distribution.

Slow and environmentally varying growth may also affect masses
inferred from standard non-accreting tracks. Even for a fixed growth
law, stars of substantially different masses can occupy similar positions in the Kiel diagram after they have finished accreting, while the equal-current-mass curves in Fig.~\ref{fig:birthline} show that stars of the same current mass occupy different positions for different formation times. An environmental spread in $\tau_0$ would further increase this non-uniqueness. This effect may contribute to the long-standing discrepancy between evolutionary and spectroscopic masses \citep[e.g.][]{Berlanas2025}, although quantifying it will require population modelling in both the Kiel and Hertzsprung--Russell diagrams.

Prolonged and environmentally varying accretion histories may also
affect subsequent evolutionary phases, with potential implications
for the empirical terminal-age main sequence
\citep{deBurgos2025,Sciarini2026} and for the long-standing difficulty
of reproducing the observed blue-supergiant population
\citep{Schootemeijer2019,Sciarini2026}. Quantifying these effects requires following populations of accreting models through later
main-sequence and post-main-sequence evolution and is left for future work.

One possible contribution to the observed ZAMS deficit is age-dependent optical selection, with the youngest massive stars
preferentially absent from spectroscopic samples because their natal environments, or even their parent molecular clouds as a whole, remain strongly obscured. Such a population-level bias cannot be excluded by
the present analysis. However, the occurrence of a similar deficit in
the Magellanic Clouds argues against an explanation tied specifically to Galactic line-of-sight extinction
\citep{Massey1995b,SabinSanjulian2017,Schootemeijer2021}. Moreover, \citet{Schootemeijer2021} examined a broad range of alternative explanations for the corresponding dearth of young massive stars in the SMC, including observational biases, uncertainties in stellar evolution, binarity, the IMF, and recent variations in the star formation rate, without finding a satisfactory explanation of both the age and luminosity distributions. Prolonged embedding remained possible, but would require a much larger population of optically hidden massive stars than had been directly identified.

The possibility that massive stars begin core-hydrogen burning while still accreting, so that their birthline is displaced from the standard ZAMS, was already demonstrated by
\citet{BeechMitalas1994} and \citet{BernasconiMaeder1996}.
More recently, \citet{Holgado2020} proposed that the observed hot-side boundary of Galactic O stars could itself trace such an accretion-modified birthline, requiring accretion to persist for an appreciable fraction of the main-sequence lifetime.
\citet{Schootemeijer2021} noted that this interpretation would require a large population of hidden massive stars still undergoing accretion. The luminosity-function analysis of
\citet{PadoanGieles2026} provides the missing demographic counterpart: the embedded ionizing population traced by compact \hii\ regions is consistent with stars that are still growing, rather than with fully formed stars simply remaining hidden after accretion has ceased. The latter possibility is also disfavoured
by cloud-scale calculations in which the natal material surrounding
individual stars is cleared rapidly \citep{Wainer+26}. In this interpretation, the compact-\hii-region lifetime problem (too many compact \hii regions relative to optically revealed OB stars) and the O-star ZAMS deficit (too few O stars near the standard ZAMS) are two observational clocks of the same extended massive-star growth phase, providing a common explanation for both long-standing discrepancies.

Our main conclusions are as follows.

(i) The accreting models naturally reproduce both the displacement
and turnover of the hot-side envelope of the Galactic O-star distribution in the Kiel diagram, implying that the end of massive-star growth does not coincide with the standard non-accreting ZAMS.

(ii) The Kiel-envelope and luminosity-function analyses provide
compatible constraints on the same growth law but have transverse
degeneracy directions. Their combination therefore substantially
tightens the allowed parameter range, yielding projected 90\%
intervals $0.43<\alpha<0.60$ and $3.9<\tau_0/{\rm Myr}<5.0$.

(iii) The jointly preferred parameters define the effective Galactic growth law $t_{\rm form}\simeq4.3\,{\rm Myr} (m_{\rm f}/60\,\msun)^{0.5}$.
Its normalization should be interpreted as characteristic of the Galactic O-star population rather than as a universal value for individual star-forming regions. 

More fundamentally, these results blur the conventional boundary between massive-star formation and stellar evolution. Massive stars can spend a substantial fraction of their core-hydrogen-burning lifetime still assembling their mass, so their internal evolution and their interaction with the surrounding gas proceed simultaneously rather than sequentially. This coupling may affect stellar structure and later evolutionary pathways, the timing and strength of radiative and mechanical feedback, and therefore the evolution of young clusters and galaxies. It may also be important in problems ranging from the origin of multiple stellar populations in globular clusters to the growth of supermassive stars at high redshift. Incorporating stellar growth into models that presently begin from fully formed massive stars is therefore likely to have consequences well beyond the ZAMS deficit addressed here.

\section*{Acknowledgements}

This research was supported by the Munich Institute for Astro-, Particle
and BioPhysics (MIAPbP), which is funded by the Deutsche
Forschungsgemeinschaft (DFG, German Research Foundation) under Germany's
Excellence Strategy -- EXC-2094 -- 390783311. This research was also
supported by the International Space Science Institute (ISSI) in Bern,
through ISSI International Team project ``The Origin of Multiple
Populations in Globular Clusters'' (ISSI Team project \#25-636).

P.P. acknowledges support by the US National Science Foundation under
Grant AST-2408023. M.G. acknowledges financial support from grants
PID2024-155720NB-I00 and CEX2024-001451-M, funded by
MCIN/AEI/10.13039/501100011033.

\section*{Data Availability}
The observational data used for comparison are taken from the published catalogues cited in the text. The MESA tracks and plotting scripts are available upon request.

\bibliographystyle{mnras}
\bibliography{references}

\clearpage
\section*{SUPPLEMENTARY MATERIAL}
\appendix

\section{Envelope selection, birthline fitting, and statistical calibration}
\label{app:method}

We constrain the massive-star growth law of equation~(2) in the main
paper by comparing the predicted end-of-accretion birthlines with the
hot-side envelope of the observed Galactic O-star distribution in the
Kiel diagram (effective temperature $\teff$ versus logarithmic surface
gravity $\log g_{\rm true}$, where the subscript “true” denotes the correction for centrifugal acceleration). Each pair of growth-law parameters
$(\alpha,\tau_0)$ defines a birthline. To carry out the comparison,
we first identify and weight the stars tracing the observed envelope,
independently of the candidate birthline being fitted. We then measure
the mismatch between this envelope and each candidate birthline and
use synthetic O-star catalogues to calibrate the resulting confidence
regions.

\emph{Envelope selection.} We use the 317-star sample described in
Section~2 of the main paper. For each star $i$, we adopt the published uncertainties
$\sigma_{T,i}$ in $\teff$ and $\sigma_{g,i}$ in
$\log g_{\rm true}$, imposing lower bounds of
$1.0\,{\rm kK}$ and $0.10\,{\rm dex}$, respectively,
to allow for systematic errors. Before selecting the envelope stars, we scale the two Kiel
coordinates by the median values of their uncertainties, so that
distances in the Kiel plane are dimensionless.
Because the observed envelope is curved, identifying its boundary
stars requires defining a local outward direction towards the
sparsely populated, hot side of the distribution. The theoretical
birthlines have approximately the same overall shape as the observed
envelope and therefore provide suitable reference curves for
defining this direction. For this purpose, we use a fixed family of 15 end-of-accretion
birthlines, with $\alpha=0,0.5,1$ and
$\tau_0=2,3,4,5,6\,{\rm Myr}$, chosen independently of the
subsequent fit. Projecting the stars onto each reference birthline defines a coordinate $s$ along the curve and a perpendicular coordinate $n$, positive in the outward direction.

Along each reference birthline, we define selection intervals of width
\begin{equation}
 W_s=f_{\rm w}\,\widetilde{\sigma}_s,
 \label{eq:window}
\end{equation}
where $\widetilde{\sigma}_s$ is the median observational uncertainty
projected onto the local tangent to the birthline, and $f_{\rm w}$
is a dimensionless width factor. Within each interval, we select
the star having the largest outward $n$. We repeat this selection for 24 shifts of the interval grid and
for every reference birthline, then take a 10\%-trimmed mean of
each star's selection frequency across the reference family.
Stars with nonzero mean selection frequencies define the envelope;
their weights $w_i$ are proportional to these frequencies and
normalized to $\sum_i w_i=1$. The lower-mass branch
is used above $\teff=33\,{\rm kK}$, whereas the higher-mass branch,
including its geometric continuation, extends to $27\,{\rm kK}$.

To assess the sensitivity of the selected envelope and the inferred
growth law to the interval width, we repeat the selection and fitting
over a broad range of $f_{\rm w}$. The best-fitting $\alpha$ and
$\tau_0$ vary smoothly with $f_{\rm w}$ over $1.25$--$2.25$, but
the trends become less regular outside this range, including a
flattening of $\alpha$ at larger widths. We therefore retain
$f_{\rm w}=1.25,1.50,1.75,2.00,2.25$. Each value defines a separate
observational envelope, whose selected stars and weights remain
fixed throughout the fitting grid, and for which the statistical
calibration is performed independently. The central value
$f_{\rm w}=1.75$ defines the fiducial envelope shown in
Fig.~1 of the main paper.

\emph{Birthline fitting.} Given a candidate birthline parameterized
continuously by $s$, we measure the squared distance of each selected envelope
star to its closest point on that birthline:
\begin{equation}
 d_i^2=\min_s\left[
 \left(\frac{T_{{\rm eff},i}-\teff(s)}{\sigma_{T,i}}\right)^2
 +\left(\frac{\log g_{{\rm true},i}-\log g_{\rm true}(s)}
 {\sigma_{g,i}}\right)^2\right].
 \label{eq:distance}
\end{equation}
Here $\sigma_{T,i}$ and $\sigma_{g,i}$ are the effective individual
errors described above; no covariance between them is assumed.
The minimum is evaluated along the piecewise-linear birthline rather
than only at tabulated points. The weighted mismatch is
$Q=\sum_i w_i d_i^2$, as defined in equation~(4) of the main paper.

\emph{Statistical calibration.} Since the envelope stars are
selected as local extremes, $Q$ need not have a standard $\chi^2$
sampling distribution. 
To calibrate confidence regions, we generate mock catalogues
with known input $(\alpha,\tau_0)$ and analyze them exactly
as we analyze the observed catalogue, including envelope
selection and birthline fitting. For each $f_{\rm w}$, we project the 317 observed stars onto the
corresponding best-fitting birthline. Their coordinates $s$ along
the birthline and their signed normal offsets $n$ from that birthline
form an empirical template of the observed distribution. 
We then use each template to generate mock catalogues for
different input growth-law parameters $(\alpha,\tau_0)$.
To construct each mock catalogue, we randomly draw 317 stars from the template, each with equal probability, and transfer them to the corresponding input birthline, preserving their fractional positions along the curve and their signed normal offsets. The same
star may be selected more than once, while others may not be
selected at all. No additional Gaussian noise is added, because
the resampled offsets already include observational scatter and
any physical scatter present in the sample.

For each $f_{\rm w}$, we generate mock catalogues using
input birthlines spanning $\alpha=0.1$--$0.9$ in steps of
$0.1$ and $\tau_0=3.0$--$5.5\,{\rm Myr}$ in steps of
$0.5\,{\rm Myr}$.
We generate 100 independent mock catalogues per input
birthline and fit each one over the full dense
$(\alpha,\tau_0)$ grid, and record
\begin{equation}
 \Delta Q_{\rm true}=Q(\boldsymbol{\theta}_{\rm true})-Q_{\min},
 \qquad\boldsymbol{\theta}=(\alpha,\tau_0),
 \label{eq:dqtrue}
\end{equation}
where $Q_{\min}$ is the minimum obtained by refitting that catalogue.
The observed quantity is defined analogously as
$\Delta Q_{\rm obs}(\boldsymbol{\theta})
=Q_{\rm obs}(\boldsymbol{\theta})-Q_{{\rm obs},\min}$.
For each $f_{\rm w}$ we pool mocks with input parameters within
$|\Delta\alpha|\leq0.40$ and
$|\Delta\tau_0|\leq1.5\,{\rm Myr}$ of the corresponding observed
best fit. The 68\%, 90\% and 95\% quantiles of the pooled
$\Delta Q_{\rm true}$ values define the thresholds for the observed
$\Delta Q$ surface. Finally, the reported Kiel-envelope confidence
regions are obtained by taking the union of the five regions calibrated separately for each adopted value of $f_{\rm w}$. The best-fitting parameters quoted in the main paper refer instead
to the fiducial selection, $f_{\rm w}=1.75$.

\section{Combining the Kiel-envelope and luminosity-function constraints}
\label{app:joint}

The Kiel-envelope analysis and the Galactic OB-star and compact
H\,{\sc ii}-region luminosity functions constrain the same growth-law
parameters through different observables. We combine their
constraints in the $(\alpha,\tau_0)$ plane, retaining the variation
with $f_{\rm w}$ described in Appendix~\ref{app:method}.
We calculate a $p$-value for each analysis at every point in this
plane, combine the two $p$-values for each $f_{\rm w}$, and take the
union of the resulting confidence regions over the five values of
$f_{\rm w}$.

For each $f_{\rm w}$, the Kiel-envelope calibration provides an
empirical distribution of $\Delta Q_{\rm true}$ from the pooled
mock catalogues described in Appendix~\ref{app:method}.
At each trial $(\alpha,\tau_0)$, we compare the observed
$\Delta Q_{\rm obs}(\alpha,\tau_0)$ with this distribution.
The Kiel $p$-value,
$p_{{\rm Kiel},f_{\rm w}}(\alpha,\tau_0)$, is the corresponding
upper-tail probability, that is the probability of obtaining a
$\Delta Q_{\rm true}$ at least as large as the observed value.
For the luminosity-function analysis of \citet{PadoanGieles2026},
we use the profiled $\Delta\chi^2_{\rm LF}(\alpha,\tau_0)$,
obtained by optimizing the three parameters describing the adopted stellar initial mass function at each $(\alpha,\tau_0)$. Under the usual $\chi^2$ approximation for two parameters
of interest, the corresponding $p$-value is
\begin{equation}
 p_{\rm LF}(\alpha,\tau_0)=
 \exp\left[-\tfrac12\Delta\chi^2_{\rm LF}(\alpha,\tau_0)\right].
 \label{eq:pLF}
\end{equation}
This field is interpolated onto the birthline-fitting grid without
extrapolation.

Our primary combination uses Fisher's method. For each
$f_{\rm w}$, we calculate
\begin{equation}
 X_{{\rm F},f_{\rm w}}=
 -2\left(\ln p_{{\rm Kiel},f_{\rm w}}+\ln p_{\rm LF}\right).
 \label{eq:fisher}
\end{equation}
Under the usual independence approximation, this statistic
follows a $\chi^2$ distribution with four degrees of freedom,
giving
\begin{equation}
 p_{{\rm F},f_{\rm w}}=
 \exp\left(-\frac{X_{{\rm F},f_{\rm w}}}{2}\right)
 \left(1+\frac{X_{{\rm F},f_{\rm w}}}{2}\right).
 \label{eq:pFisher}
\end{equation}
For a nominal confidence level $C$, the region for each
$f_{\rm w}$ consists of points satisfying
$p_{{\rm F},f_{\rm w}}\geq1-C$.
We report the union of these five regions, equivalently
obtained from
$p_{\rm F}=\max_{f_{\rm w}}p_{{\rm F},f_{\rm w}}$.
The 90\% confidence intervals for $\alpha$ and $\tau_0$ quoted in the main text are obtained by projecting the
union of the five joint 90\% confidence regions onto
the respective parameter axes. The joint optimum marked in Fig.~3 of the main paper
corresponds to the fiducial $f_{\rm w}=1.75$.

\label{lastpage}
\end{document}